\documentclass[preprint, superscriptaddress]{revtex4-1}
\usepackage{graphicx}
\usepackage{amsmath}
\usepackage{amsfonts}
\usepackage{amssymb}
\usepackage{tabularx}
\usepackage{comment}
\usepackage{cleveref}
\newcommand{\kvec}{\boldsymbol{k}}
\newcommand{\qvec}{\boldsymbol{q}}
\newcommand{\lvec}{\boldsymbol{l}}

\newcommand{\nvec}{\boldsymbol{n}}

\newcommand*{\Scale}[2][4]{\scalebox{#1}{$#2$}}%

\begin{document}

\title{Role of substrate induced electron-phonon interactions in biased graphitic bilayers}

\begin{abstract}
Bilayers of graphitic materials have potential applications in field effect transistors (FETs). A potential difference applied between certain ionic bilayers made from insulating graphitic materials such as BN, ZnO and AlN could reduce gap sizes, turning them into useful semiconductors. On the other hand, opening of a small semiconducting gap occurs in graphene bilayers under applied field. The aim here is to investigate to what extent substrate induced electron-phonon interactions (EPIs) modify this gap change. We examine EPIs in several lattice configurations, using a perturbative approach. The typical effect of EPIs on the ionic bilayers is an undesirable gap widening. The size of this gap change varies considerably with lattice structure and the magnitude of the bias. When bias is larger than the non-interacting gap size, EPIs have the smallest effect on the
bandgap, especially in configurations with $AA^{\prime}$ and $AB$ structures. 
Thus careful selection of substrate, lattice configuration
and bias strength to minimise the effects of EPIs could be
important for optimising the properties of electronic devices. We use parameters related to BN in this article. In practice, the results presented here are broadly applicable to other graphitic bilayers, and are likely to be qualitatively similar in metal dichalcogenide bilayers such as MoS$_2$, which are already of high interest for their use in FETs.
\end{abstract}


\author{A. R. Davenport}
\affiliation{Department of Physical Sciences, The Open University, Walton Hall, Milton Keynes MK7 6AA, UK}

\author{J. P. Hague}
\affiliation{Department of Physical Sciences, The Open University, Walton Hall, Milton Keynes MK7 6AA, UK}
\maketitle


\section{Introduction}

\begin{figure*}

\begin{center}
\begin{tabular}{| c | *3{>{\centering\arraybackslash}p{.27\textwidth} | } }
\hline
{} & Lattice Configuration & Hamiltonian & Parameters / eV \\ 
\hline
$AA^\prime$ & 
\includegraphics[width=0.17\textwidth]{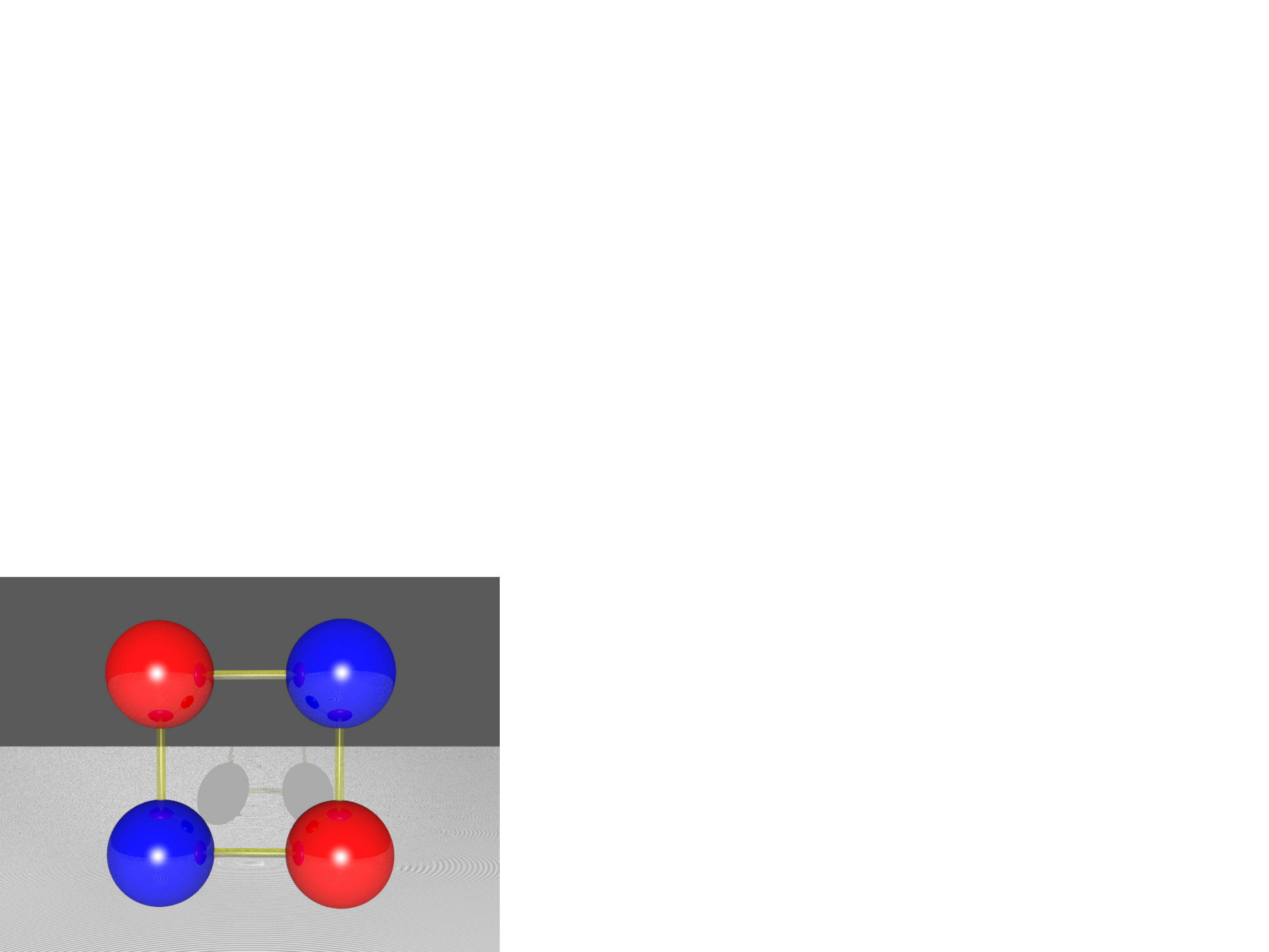}
&\vspace{-2.0cm}$ 
\Scale[1.0]{\left( \begin{array}{c c c c}
-\Delta & \Phi_{\kvec} & 0 & \gamma_1 \\ 
\Phi_{\kvec}^* & \Delta & \gamma_1 & 0 \\ 
0 & \gamma_1 & -\Delta & \Phi_{\kvec}^* \\ 
\gamma_1 & 0 & \Phi_{\kvec} & \Delta
\end{array} \right)}$ &  
\vspace{-2.0cm}$\begin{array}{c}
\gamma_0 = 2.36 \\
\gamma_1 = 0.32 \\
\Delta \approx 2.04
\end{array}$ \\
\hline
$AB$ & 
\includegraphics[width=0.17\textwidth]{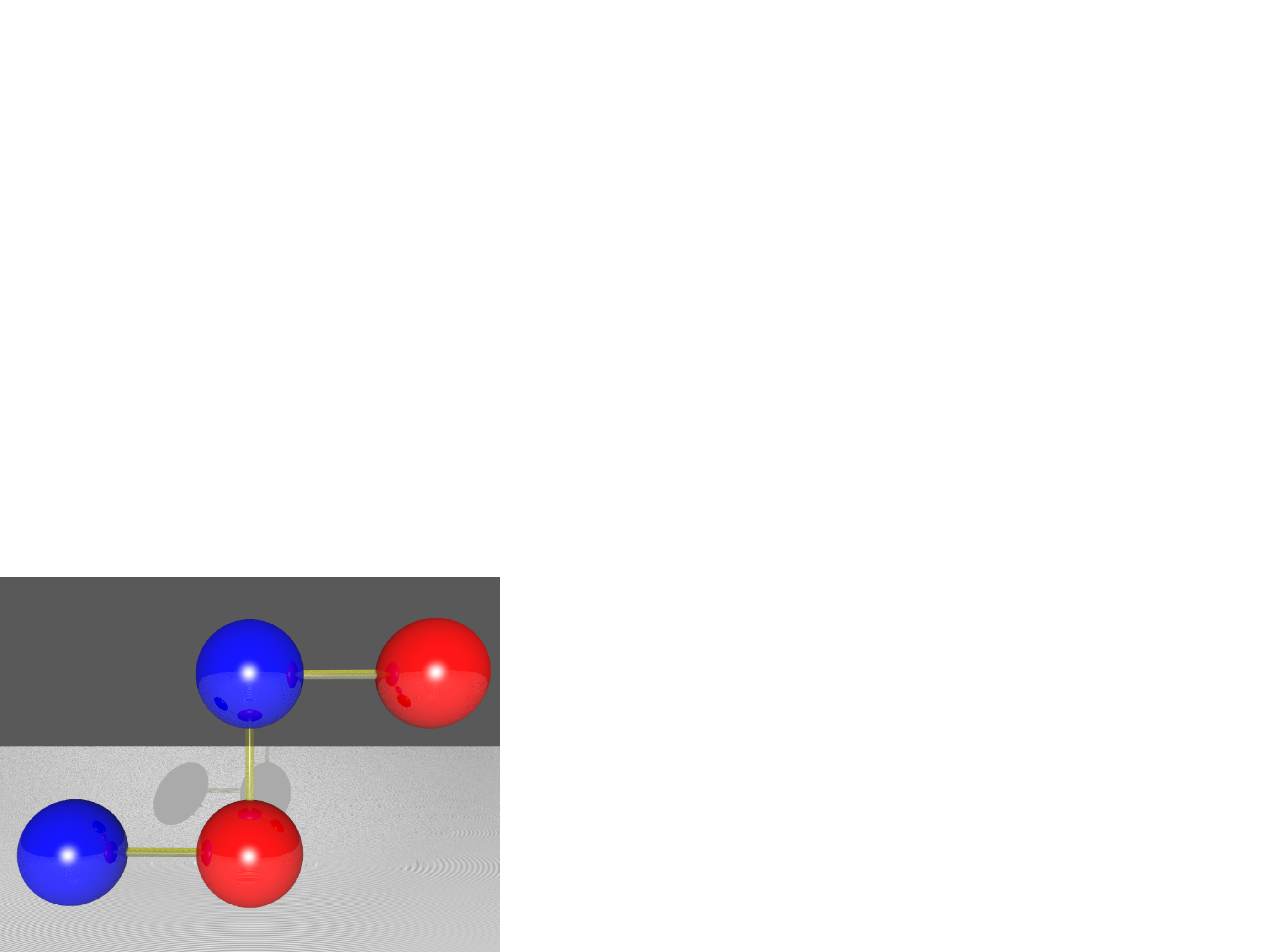}
&\vspace{-2.0cm} $ 
\Scale[1.0]{\left( \begin{array}{c c c c}
-\Delta & \Phi_{\kvec} & 0 & 0 \\ 
\Phi_{\kvec}^* & \Delta & \gamma_1 & 0 \\ 
0 & \gamma_1 & -\Delta & \Phi_{\kvec}^* \\ 
0 & 0 & \Phi_{\kvec} & \Delta
\end{array} \right)}$ & 
\vspace{-2.0cm}$\begin{array}{c}
\gamma_0 = 2.37 \\
\gamma_1 = 0.60 \\
\Delta \approx 2.08
\end{array}$ \\
\hline
$A^\prime B$ &
\includegraphics[width=0.17\textwidth]{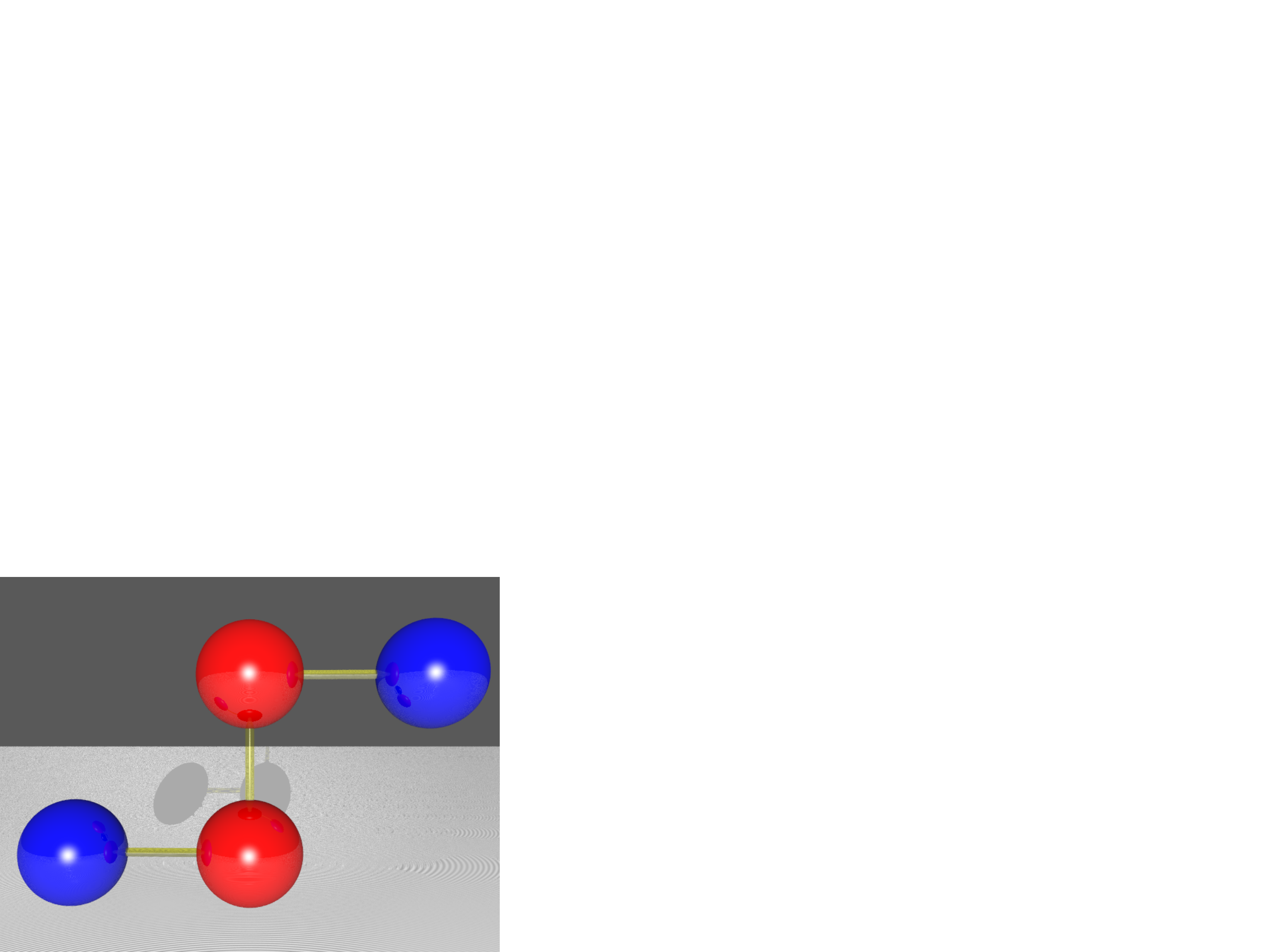} 
& \vspace{-2.0cm}$ 
\Scale[1.0]{\left( \begin{array}{c c c c}
-\Delta & \Phi_{\kvec} & 0 & 0 \\ 
\Phi_{\kvec}^* & \Delta & \gamma_1 & 0 \\ 
0 & \gamma_1 & \Delta & \Phi_{\kvec}^* \\ 
0 & 0 & \Phi_{\kvec} & -\Delta
\end{array} \right)}$  & 
\vspace{-2.0cm}$\begin{array}{c}
\gamma_0 = 2.34 \\
\gamma_1 = 0.25 \\
\Delta \approx 1.96
\end{array}$ \\
\hline
$AB^\prime$ & 
\includegraphics[width=0.17\textwidth]{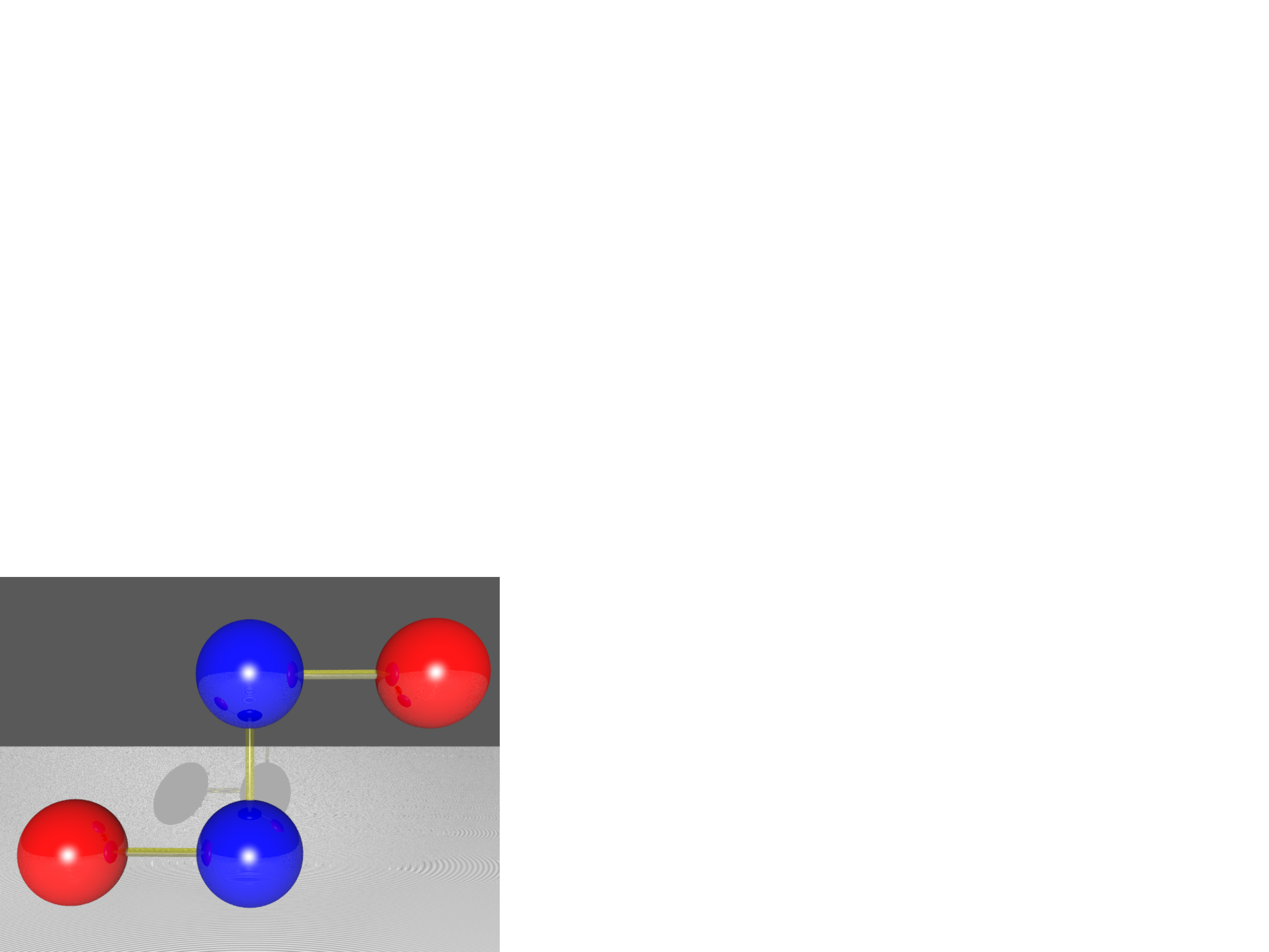}
&\vspace{-2.0cm} $ 
\Scale[1.0]{\left( \begin{array}{c c c c}
\Delta & \Phi_{\kvec} & 0 & 0 \\ 
\Phi_{\kvec}^* & -\Delta & \gamma_1 & 0 \\ 
0 & \gamma_1 & -\Delta & \Phi_{\kvec}^* \\ 
0 & 0 & \Phi_{\kvec} & \Delta
\end{array} \right)}$ & 
\vspace{-2.0cm}$\begin{array}{c}
\gamma_0 = 2.38 \\
\gamma_1 = 0.91 \\
\Delta \approx 2.16
\end{array}$ \\
\hline 
$AB$ (graphene) & 
\includegraphics[width=0.17\textwidth]{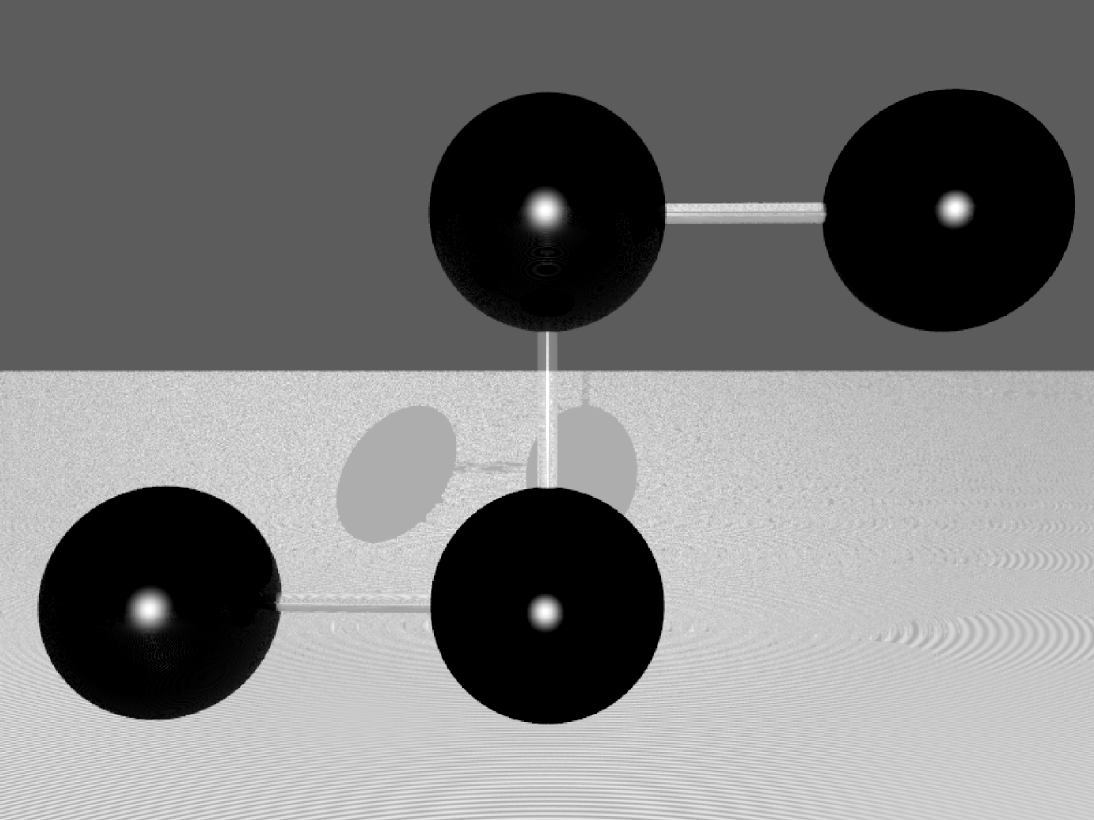}
& \vspace{-2.0cm}$ 
\Scale[1.0]{\left( \begin{array}{c c c c}
-\Delta & \Phi_{\kvec} & 0 & 0 \\ 
\Phi_{\kvec}^* & \Delta & \gamma_1 & 0 \\ 
0 & \gamma_1 & \Delta & \Phi_{\kvec}^* \\ 
0 & 0 & \Phi_{\kvec} & -\Delta
\end{array} \right)}$ & 
\vspace{-2.0cm}$\begin{array}{c}
\gamma_0 = 3.00 \\
\gamma_1 = 0.30 \\
\Delta = 0.0105
\end{array}$ \\
\hline 
\end{tabular}
\end{center}

\caption{Lattice configurations and tight-binding parameters of the bilayer boron nitride and graphene configurations. BN values are taken from
Ref. \onlinecite{Ribeiro2011}, and graphene ones mimic experiment \cite{zhang2008}. Red circles represent N, blue ones represent B and black ones represent C. The small $\Delta$ in graphene originates from the proximity of carbon atoms in different layers.} \label{TBtable}
\end{figure*}

The discovery and manufacture of graphene has given a significant
boost to research into two-dimensional materials \cite{Novoselov2005},
often with the aim of integrating them into electronic
devices.\cite{Teweldebrhan2010} These new two-dimensional materials
are robust, with high crystal qualities. However, with the exception
of graphene, many of these materials are relatively
unexplored.\cite{Ribeiro2011} For example, hexagonal boron nitride
(h-BN) is a low dimensional material of high interest, because of its
many similarities to graphene; a honeycomb lattice structure and a
similar bond length, although the bonds between boron and nitrogen
atoms have a high degree of ionicity in comparison to the covalent
bonds in graphene.\cite{Topsakal2009} This ionicity leads to a band
gap, and may also be found in other ionic graphitic bilayers such as ZnO,
GaN, AlN, BeO and MgO.  Another subtle difference is that graphene has
two possible configurations of atoms, whereas ionic graphitic bilayers
have four possible stable bilayer configurations due to the two
distinct atom types in the bilayer (see \Cref{TBtable}). All four
configurations display differing characteristics that could be
desirable, with a range of tight binding parameters and energy gaps
(for example all of the BN configurations have gaps of order $4$eV).\cite{Warner2010}


Of particular interest for the current article is the prediction by
Tang {\it et al.} that applying a bias across bilayer h-BN can close
the BN gap.\cite{tang2012} Such an effect is not limited to bilayers
of BN, and is also predicted to occur in bilayer MoS$_2$, which has
attracted a great deal of interest for use in field effect
transistors.\cite{liu2012} Given the similarity of graphitic phases of
compounds such as ZnO, GaN, AlN, BeO and MgO to BN, and similarity of
MoS$_2$ to a wide variety of other metal
dichalcogenides,\cite{tang2013} such band closing effects could be
widespread.

In this article, we will use parameters relating to BN, although the
other ionic graphitic materials have similar properties. We note that other
theoretical and experimental approaches to manipulate h-BN band
structure have been similar to those used to make gaps in graphene (and it is likely that similar approaches would be applicable to other ionic graphitic bilayers)
\cite{Lin2012}. Examples include the alteration of the structural geometry of
BN, introduction of impurities by replacing B or N atoms or by adding
adatoms, creating h-BN nanotubes, amplifying ionicity with substrate
mediated electron-phonon interactions or by forming
nanoribbons.\cite{Golberg2007, Wang2013, Hague2012} An interesting
effect has been seen when h-BN bilayers are functionalised with
hydrogen; if both layers are fully saturated the size of the band gap
is reduced and the gap changes from direct to indirect.\cite{Zhou2010,
  Wang2010}



Given the wide range of graphitic materials with gap closure under bias, the aim of this paper is to investigate the effect of electron-phonon interactions on those materials, especially on the band closing effect - does it enhance or reduce the effect, and are there situations where the effects of EPI can be minimised? In this paper we
focus on modifications to the band gap using parameters related to biased bilayer
BN \cite{zhai2013} caused by EPIs between the
bilayer and a substrate \cite{Hague2012}. With the exception of graphene, we are not aware of other studies of EPIs in biased graphitic bilayers. The model may also be considered as an approximation to in-plane EPIs. We also consider the effect of EPIs on biased bilayer graphene. The paper is structured as
follows; in Sec. \ref{Sec:Model} we introduce our model and perform a
Green's function analysis to determine the low order contributions to
the self-energy, and therefore to derive self-consistent
equations. Section \ref{Sec:Results} presents numerical solutions to
the Green's function, which are used to determine the gap
enhancement. We examine the effects of EPIs on bilayers of graphitic materials for a range of structures that couple to substrates and sandwiches of polar materials. Finally, we summarise, discuss limitations and further work and make conclusions in
Sec. \ref{Sec:Conclusion}.

\section{Model}\label{Sec:Model}

We model hexagonal boron nitride bilayers using a tight-binding
approach, with a bias applied normal to the bilayer surface. We
introduce an additional electron-phonon term describing the
interactions between electrons in the bilayer and phonons in the substrate (or superstrate if the bilayer is sandwiched). There are several possible forms of the
electron-phonon interaction. The two main classes are
those where (a) local electron density couples to phonon modes
directly, such as the Holstein and extended Holstein (Fr\"ohlich) models \cite{holstein1959,frohlich1954,alexandrov2002}, and (b)
significant distortions of the lattice modify hopping integrals and therefore lead
to interaction with phonon modes when electrons move, as is the case in the Su-Schrieffer-Heeger model of polyacetylene and other polymers \cite{su1980}. The deformability of the system is critical for determining which of
types (a) and (b) dominate: If materials are not very flexible or
compressible, then type (a) dominates. For systems which are flexible,
such as polymers, then interactions of type (b) are more
typical. Freestanding and suspended graphene are also flexible, and
electrons in the plane can interact with out of plane flexural modes through coupling of the hopping to a local vector potential representing the local deformation \cite{Graphene101} (which is in essence a 2D extension of the SSH interaction), whereas the presence of substrates makes the graphene planes rigid and leads to suppression of flexural modes \cite{mayorov2011}. The graphitic bilayer systems considered in this paper exist on substrates, and therefore the electron-phonon interactions are expected to be between electrons in the plane and phonons in the substrate, where the
extended Holstein (or Fr\"ohlich) forms dominate. The extended Holstein form is known to be the dominant type of interaction in layered materials \cite{alexandrov2002}, including 2D materials such as graphene on substrate systems (see e.g. \cite{hwang2013}), and this form has been measured directly between carbon nanotubes and SiO$_2$ substrates \cite{steiner2009}. We expect this to be the case for all graphitic systems on substrates (or in sandwiches) such as those studied here.

Extended Holstein and local
Holstein forms of interaction have qualitatively similar
properties.  On the mean-field
level, Holstein and Fr\"ohlich interactions are identical \cite{Hague2014} due to averaging of the interaction across the Brillouin zone \cite{maier2005}. Therefore,
the form of the interaction is taken to be of the
local, Holstein, form \footnote{Quantitative study of the extended Holstein form requires us
to go beyond the mean-field level by using approaches such as the
dynamical cluster approximation \cite{Hague2014}},
\small
\begin{align}
H = &-\gamma_0 \sum_{\langle n,n^\prime \rangle u\sigma}
(\alpha_{n u \sigma}^\dagger \beta_{n^\prime u \sigma} + 
\beta_{n^\prime u \sigma}^\dagger \alpha_{n u \sigma}) \nonumber\\
&-\gamma_1 \sum_{\sigma n} (X^\dagger_{n1\sigma} Y_{n^\prime2\sigma} + 
Y^\dagger_{n^\prime2\sigma} X_{n1\sigma}) 
- \sum_{n u \sigma} g_{u} n_{n u \sigma}x_n \\&+ \sum_m \hbar\Omega(N_m + \frac{1}{2})
+ \sum_{n u \sigma} \Delta_{n u} n_{n u \sigma} + \sum_{nu\sigma}
V_u n_{nu\sigma} \label{EQN::Hamiltonian} \nonumber
\end{align}
\normalsize
Here, $\gamma_0$ and $\gamma_1$
represent the hopping integrals for intra-layer and inter-layer
hopping respectively. $\alpha^{\dagger}_{i\sigma}$ creates an electron on a boron site with a spin $\sigma$ and lattice vector $i$, and the operator $\beta$ annihilates an electron on the nitrogen sublattice. The index $u$ indicates the layer in which the site sits, and we use the symbols $\bigtriangledown$ and $\bigtriangleup$ for the layers respectively closest and furthest from the substrate. $V_{\bigtriangleup}=+V$ and $V_{\bigtriangledown}=-V$. $X^{\dagger}$ and $Y^{\dagger}$ are either $\alpha^{\dagger}$ or $\beta^{\dagger}$ depending on the specified lattice configuration. Inter-layer terms are taken into account when
an atom $X$ sits directly below an atom $Y$ in the lattice structure,
where $X$ and $Y$ can represent the boron and / or nitrogen atoms (or other atoms if alternative graphitic bilayers are to be considered). $g_{u}$ determines the magnitude of interaction
between electrons and ions on sites at position $\nvec$, and can be
related to the dimensionless electron-phonon coupling
$\lambda_{u}=g_{u}^2/2M\Omega^2 \gamma_0$. $M$ is the ion mass, $\Omega$ is the phonon 
frequency, $n$ the electron number operator, $N_m$ the phonon number
operator and $x_n$ the ion displacement. Finally, $\Delta_{nu}$ introduces the 
atomic ionicity of each site, and $V$ is the magnitude of the potential at each plane, such that the total potential difference between the planes is $2V$.

Low order perturbation theory can be used to solve this Hamiltonian
when phonon frequencies are low and electron-phonon coupling constants
are weak. We construct the full Green's function of the system using
Dyson's equation,
$\boldsymbol{G}^{-1}(\kvec,i\omega_n)=\boldsymbol{G}_0^{-1}(\kvec,i\omega_n)-\boldsymbol{\Sigma}(i\omega_n)$,
substituted with the following form for the self-energy,
\begin{gather}
\Sigma_{jj}(i\omega_n) \approx i\omega_n (1-Z_j) + \delta_{j}
\end{gather}
The number of unique sites dictates the minimum number of functions
that are used to describe the system. In the following, we start by
examining the general case with four modified potentials, $\delta_1$,
$\delta_2$, $\delta_3$ and $\delta_4$ which represent the effects of
interactions on the four sites of the unit cell. In addition, the
respective quasi-particle weights $Z_1$, $Z_2$, $Z_3$ and $Z_4$ are
included. Both $\delta$ and $Z$ are real functions of the Matsubara frequency for
fermions, $\omega_n = 2\pi k_BT(n+1/2)$. Off diagonal terms in the
self-energy are zero in the low order perturbation theory considered
here and terms are completely momentum independent. The
non-interacting Green's function of the system can be found from,
$\boldsymbol{G}_0^{-1}(k,i\omega_n)=[\boldsymbol{I}i\omega_n-(\boldsymbol{H}+\boldsymbol{V})]$. Where $\boldsymbol{H}$ is defined in Fig. \ref{TBtable} and $\boldsymbol{V}$
represents the applied bias,
\begin{equation}
\boldsymbol{V} = \left(\begin{array}{cccc}V & 0 & 0 & 0 \\ 0 & V & 0 & 0 \\ 0 & 0 & -V & 0 \\ 0 & 0 & 0 & -V\end{array}\right)
\end{equation}

We invert Dyson's equation and then place it into
the lowest order contribution to the self-energy, which is,
\begin{eqnarray}
\Sigma_{ij}(\kvec,i\omega_n) = -T\gamma_0\lambda_{ij}\sum_{i\omega_s}  \int \frac{\mathrm{d}^2\qvec}{V_{BZ}}\boldsymbol{G}_{ij}(\kvec-\qvec,i\omega_{n-s})\\
\times  \left[2d_0(\qvec,\omega_{s=0}) - d_0(\qvec,\omega_s)\right],\nonumber
\label{EqnLowOrderPert}
\end{eqnarray}
\normalsize
In this equation, the non-interacting phonon propagator is
$d_0(i\omega_s)=\delta_{ij}\Omega^2/(\Omega^2-\omega_s^2)$, and
the Matsubara frequencies for bosons are, $\omega_s=2\pi k_B T s$. $\lambda_{ij}$ is defined from,
\begin{equation}
\boldsymbol{\lambda} = \left(\begin{array}{cccc}\lambda_{\bigtriangledown} & 0 & 0 & 0 \\ 0 & \lambda_{\bigtriangledown} & 0 & 0 \\ 0 & 0 & \lambda_{\bigtriangleup} & 0 \\ 0 & 0 & 0 & \lambda_{\bigtriangleup}\end{array}\right)
\end{equation}
For the case of substrate and superstrate, $\lambda_{\bigtriangledown}=\lambda_{\bigtriangleup}=\lambda$. For the case where there is only a substrate, but no covering superstrate, $\lambda_{\bigtriangleup}$ would have a tiny value on the order of 2-3\% of $\lambda_{\bigtriangledown}$ (assuming the upper layer is about twice as far from the substrate as the lower layer, since the coupling, $\lambda\sim g^2$ and $g$ goes like $1/r^3$), so we take the coupling to be $\lambda_{\bigtriangledown}=\lambda$, and $\lambda_{\bigtriangleup}=0$.

Thus, we obtain four sets of simultaneous equations for each
configuration, that describe how the effective potential, $\delta$ and
quasi-particle weight, $Z$, change with our input parameters;
temperature, phonon frequency, on-site potential and electron-phonon
coupling constant,
\begin{widetext}
\small
\begin{eqnarray}
\delta_1  +  i \omega_n (1-Z_1)  & = & \gamma_0\lambda_{\bigtriangledown} k_BT\sum_f [2d_0(i\omega_{s=0})-d_0(i\omega_s)]\int \mathrm{d}\varepsilon \frac{  D(\varepsilon)\, (\varepsilon^2\Pi_2 + \Pi_4( \gamma_1^2-\Pi_2\Pi_3))}{(-\varepsilon^4 + \Pi_1\Pi_4( \gamma_1^2-\Pi_2\Pi_3)) + \varepsilon^2( \Pi_1\Pi_2 + \Pi_3\Pi_4)}
\label{Eqn:ABDelta1} \\
\delta_2 + i \omega_n (1-Z_2) & = & \gamma_0\lambda_{\bigtriangledown} k_BT\sum_f [2d_0(i\omega_{s=0})-d_0(i\omega_s)]\int \mathrm{d}\varepsilon \frac{ D(\varepsilon) \,(\varepsilon^2\Pi_1 + \Pi_3(\Pi_1\Pi_4))}{(-\varepsilon^4 + \Pi_1\Pi_4( \gamma_1^2-\Pi_2\Pi_3)) + \varepsilon^2( \Pi_1\Pi_2 + \Pi_3\Pi_4)}
\label{Eqn:ABDelta2} \\
\delta_3 + i \omega_n (1-Z_3) & = & \gamma_0\lambda_{\bigtriangleup} k_BT\sum_f [2d_0(i\omega_{s=0})-d_0(i\omega_s)] \int \mathrm{d}\varepsilon \frac{  D(\varepsilon)\,(\varepsilon^2\Pi_4 + \Pi_2(\Pi_4\Pi_1))}{(-\varepsilon^4 + \Pi_1\Pi_4( \gamma_1^2-\Pi_2\Pi_3)) + \varepsilon^2( \Pi_1\Pi_2 + \Pi_3\Pi_4)}
\label{Eqn:ABDelta3} \\
\delta_4 + i \omega_n (1-Z_4) & = & \gamma_0\lambda_{\bigtriangleup} k_BT\sum_f [2d_0(i\omega_{s=0})-d_0(i\omega_s)] \int \mathrm{d}\varepsilon\frac{  D(\varepsilon) \,(\varepsilon^2\Pi_3 + \Pi_1( \gamma_1^2-\Pi_3\Pi_2))}{(-\varepsilon^4 + \Pi_1\Pi_4( \gamma_1^2-\Pi_2\Pi_3)) + \varepsilon^2( \Pi_1\Pi_2 + \Pi_3\Pi_4)}
\label{Eqn:ABDelta4} 
\end{eqnarray}
\normalsize

which are valid for the the $AB^\prime$, $A^{\prime}B$ and $AB$ configurations. In the case of the $AA^{\prime}$ stacked
configuration, atoms in different planes sit above or below every site, so equations (\ref{Eqn:ABDelta2}) and (\ref{Eqn:ABDelta3}) are replaced with a different form,
\small
\begin{eqnarray}
\delta_2 + i \omega_n (1-Z_2) & = & \gamma_0\lambda_{\bigtriangledown} k_BT\sum_f [2d_0(i\omega_{s=0})-d_0(i\omega_s)] \int \mathrm{d}\varepsilon \frac{  D(\varepsilon)\,(\varepsilon^2\Pi_1 + \Pi_3( \gamma_1^2-\Pi_1\Pi_4))}{(-\varepsilon^4 + \Pi_1\Pi_4( \gamma_1^2-\Pi_2\Pi_3)) + \varepsilon^2( \Pi_1\Pi_2 + \Pi_3\Pi_4)}
\label{Eqn:AADelta1} \\
\delta_3 + i \omega_n (1-Z_3)&  = & \gamma_0\lambda_{\bigtriangleup} k_BT\sum_f [2d_0(i\omega_{s=0})-d_0(i\omega_s)]\int \mathrm{d}\varepsilon\frac{  D(\varepsilon) \,(\varepsilon^2\Pi_4 + \Pi_2( \gamma_1^2-\Pi_4\Pi_1))}{(-\varepsilon^4 + \Pi_1\Pi_4( \gamma_1^2-\Pi_2\Pi_3)) + \varepsilon^2( \Pi_1\Pi_2 + \Pi_3\Pi_4)}
\label{Eqn:AADelta2} 
\end{eqnarray}
\normalsize
\end{widetext}
where, $f=n-s$. In the self-consistent equations, all terms are
momentum independent except for terms of the form,
$\Phi_{\kvec}\Phi^{*}_{\kvec}$ and $(\Phi_{\kvec}\Phi^{*}_{\kvec})^2$,
where $\Phi_{\kvec}=\gamma_0\sum_{\lvec}e^{-i\kvec \cdot\lvec}$ and
the sum is over the nearest neighbour vectors within the same layer, $\lvec$. The products $\Phi\Phi^{*}$ have the
same form as the squared dispersion in the monolayer case, and
therefore as a mathematical tool the sum over momenta can be replaced
by an integral over the monolayer density of states,
$D(\varepsilon)$. This is mathematically identical to the
identification of the bilayer dispersion and replacement of the
momentum sum with the bilayer density of states. The rewriting in
terms of the monolayer DOS leads to the slight advantage of a
straightforward analytical form for the DOS at all energies, although
in the following we will use a linear approximation, $D(\varepsilon)=|\varepsilon|/\pi\gamma_{0}^2\sqrt{3}$, for $|\epsilon|<\gamma_{0}\pi^{1/2}3^{1/4}$.\cite{Graphene101}

Each configuration has separate definitions of $\Pi_x$, although all have the
general form, $\Pi_x(i\omega_n) = \delta_x \pm \Delta \pm V +
i\omega_nZ_x$, with sign changes relating to the properties of site
$x$. All $\delta_{x}$ and $Z_{x}$ are all taken to be real so each
equation is solvable by separating real and imaginary parts. Summing
over all Matsubara frequencies, truncated at sufficiently large
$\omega_n$ to ensure convergence, the equations can be solved
self-consistently ($\omega_{n_{max}}=240\gamma_0$). Calculations for
each bilayer configuration were performed separately according to
their tight binding parameters (seen in \Cref{TBtable}), in addition
to the potential bias placed perpendicularly over the two planes, and the electron-phonon interaction. For comparison, we also make computations for bilayer graphene.

\section{Results}\label{Sec:Results}

\begin{figure}
  \includegraphics[width = 0.5\textwidth]{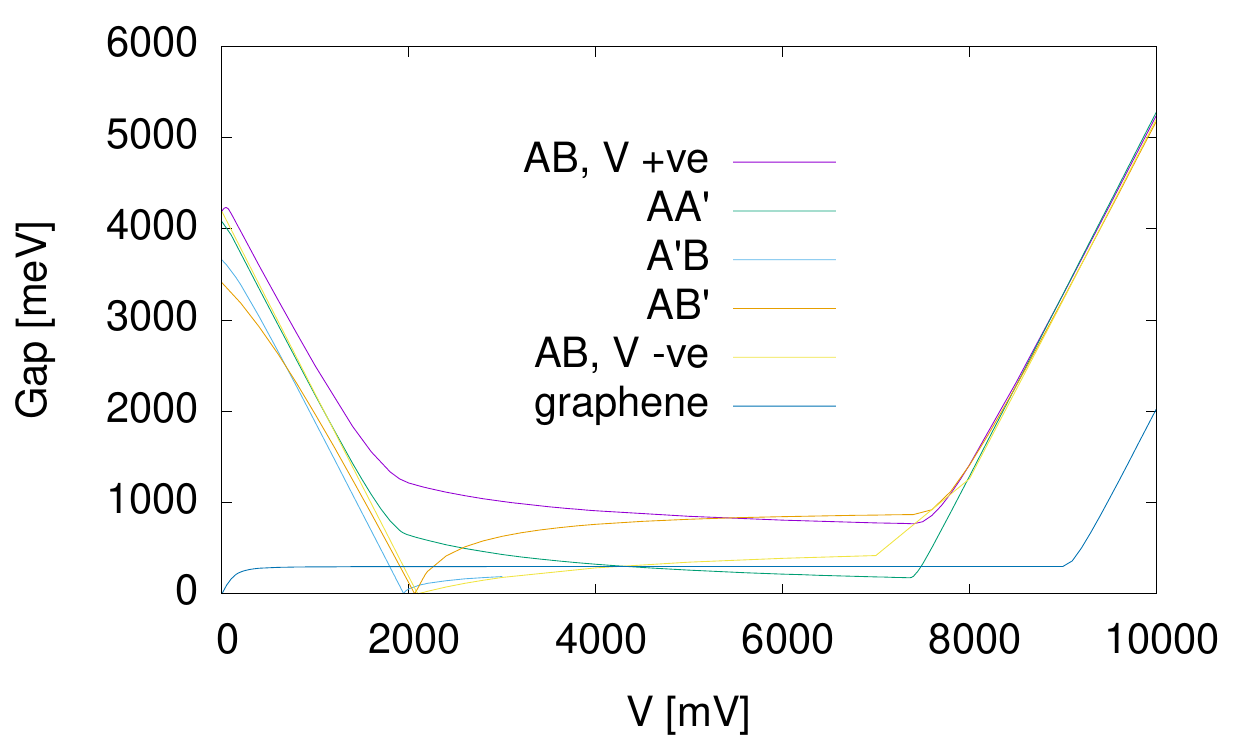}
  \caption{Band gap evolution for the alternative configurations of BN and graphene as the potential difference perpendicular to the planes is changed, when interactions are not present. There are 3 main forms of the response of the gap to potential, including regions where the gap does not change rapidly with potential, and specific values of $V$ where the gap is reduced to zero.}
\label{fig:gapvsv}
\end{figure}

Tight binding parameters for each of the lattice configurations can be
found in Figure \ref{TBtable} where they have been extracted from
Ref. \onlinecite{Ribeiro2011}. All intra-layer hopping parameters in
the different h-BN configurations are very similar (within $1\%$ of
$2.36$eV). On the other hand, inter-layer hopping parameters are
highly dependent on stacking configurations, ranging from $\gamma_1 =
0.25$eV to $\gamma_1 = 0.91$eV. The band gap for all configurations is
approximately $4$eV before electron-phonon interactions and the
interlayer potential are switched on. The interlayer potential modifies the band gap, with different changes in gap size for the alternative configurations of BN and graphene, and these can be seen in Fig. \ref{fig:gapvsv}. There are 3 main forms of the response of the gap to potential. In bilayer graphene, the gap rapidly increases with potential until it reaches a plateau of around 300meV. This plateau is very wide, persisting up to 9V ($3\gamma_{0}$). The $A'B$, $AB'$ and (when negative potential difference is applied) $AB$ forms of BN undergo gap reduction until $V\approx\Delta$. After this, the gaps slowly rise again, not quite reaching a plateau until a point of inflection around $V=3\gamma_{0}$ where the gradient starts to increase rapidly. Finally, the $AA'$ stacked form and the $AB$ form with positive potential difference have a rapid gap decrease to a point of inflection around $V=\Delta$, and then the gap remains only weakly changed up to $V=3\gamma_{0}$, where there is a minimum followed by a rapid increase of the gap with $V$. The difference between the application of positive and negative potential in the $AB$ case relates to the asymmetry between B and N atoms on the central sites of the unit cell. No such asymmetry exists for the other cases.

In the following, temperature is set at $k_BT=0.01\gamma_0$, which is $\approx 24$meV for BN and $\approx 30$meV for graphene,
corresponding to a range between $266$K and $278$K dependent on the value of $\gamma_0$, although the results are essentially temperature independent around room temperature (we checked for $k_BT=48$meV or $540$K, obtaining essentially identical results). Phonon energies of
$\hbar\Omega = 0.02\gamma_0$ were used in the
calculations (corresponding to $\hbar\Omega \approx 48$meV for BN, depending on the differences in $\gamma_0$, and $\hbar\Omega \approx 60$meV for graphene). Computations were also made for a higher value of $\hbar\Omega = 0.06\gamma_0$, with no qualitative changes to the results.

Equations \ref{Eqn:ABDelta1}-\ref{Eqn:AADelta2} were solved numerically with a linear approximation to the the density of states to lower the computational costs. Calculations were conducted at several points of interest in boron nitride bilayers; the point at which the bias produces an electron band gap of $1$eV (similar in size to that found in silicon) and the point at which the gap size becomes zero in certain configurations (found to be close to $V=\Delta$). Specific values of $V$ used for the different BN structures can be found in Table \ref{Tab:BNBias}. Calculations were also carried out for $V=\gamma_0$ and $V=2\gamma_0$, where the gap varies less quickly on change of $V$.

\begin{table}[!ht]
\begin{center}
\begin{tabular}{|c|c|c|}
\hline
Configuration & V (1eV gap) & V (zero gap \\ 
& &  / point of inflection) \\\hline
$AA^\prime$ & 1.655eV & 2.04eV \\ \hline
$AB$ (+ve) & 3.114eV & $+ 2.08$eV \\ \hline
$AB$ (-ve) & -1.595eV & $- 2.08$eV \\ \hline
$A^\prime B$ & $1.445$eV & 1.96eV \\ \hline
$AB^\prime$ & $1.536$eV & 2.06eV \\ \hline
\end{tabular}
\caption{Required potential bias between h-BN sheets to produce a band-gap of 1eV, and $V$ corresponding to zero gap or point of inflection depending on configuration. N.B. There are differences between the gaps at +ve and -ve $V$ for the AB configuration.}
\label{Tab:BNBias} 
\end{center}
\end{table}

\begin{figure*}[!ht]
\includegraphics[width=0.85\textwidth]{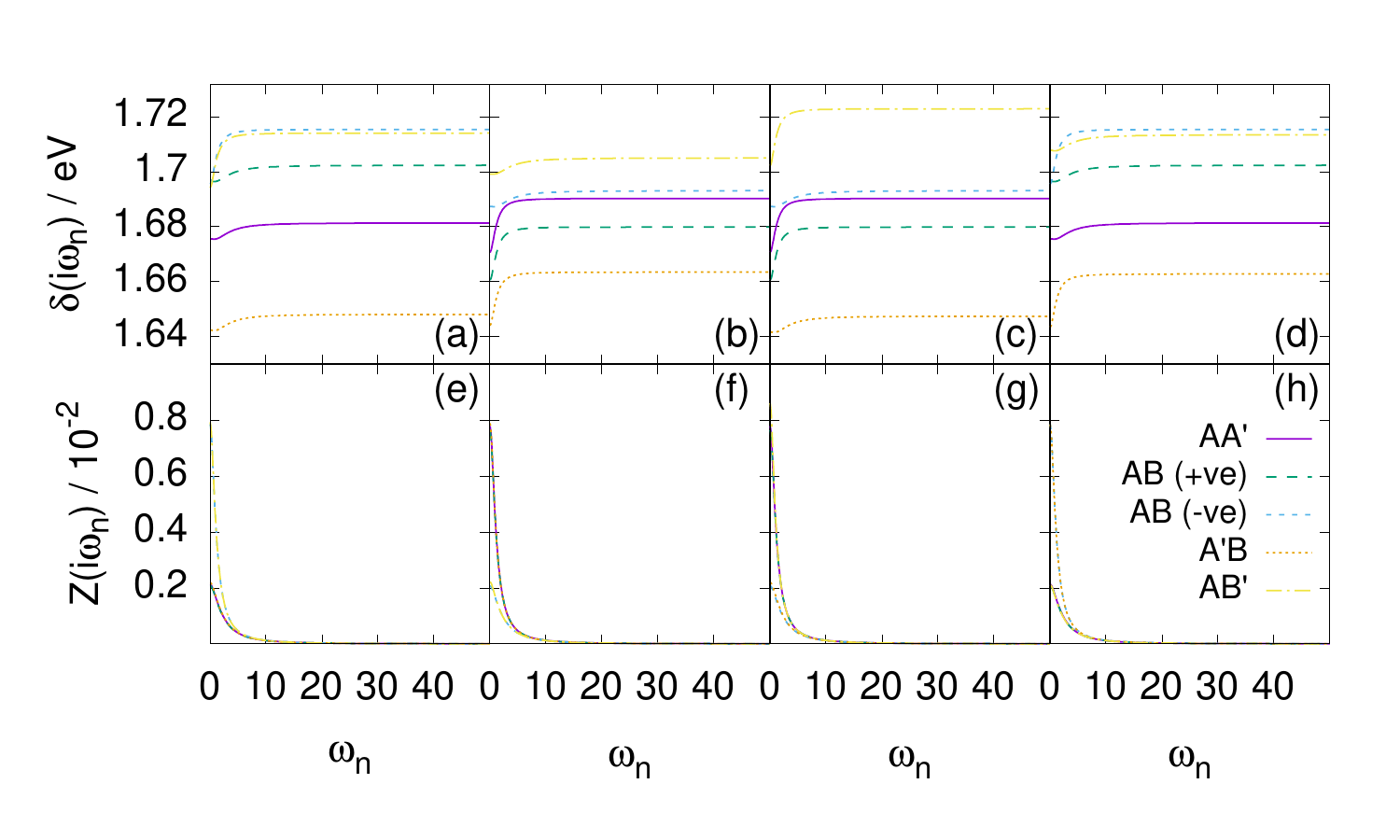}
\caption[Matsubara frequency dependence for bilayer boron nitride on-site potentials]{Panels (a)-(d). Matsubara frequency dependence of the magnitude of the bilayer boron nitride on-site potentials (a) $\delta_1$ to (d) $\delta_4$ for an electron-phonon coupling $\lambda=1$, at the point of inflection / zero gap around $V\approx\Delta$. The bilayer is completely sandwiched in this case, but the forms of the functions are similar for coupling to substrate only. Each panel shows a different sub-lattice site. Panels (e)-(h) show values for the associated quasi-particle weights $Z_{1}$ through to $Z_{4}$.}
\label{Fig:BNmatsubara}
\end{figure*}

\begin{figure*}
\centering
\includegraphics[width=0.85\textwidth]{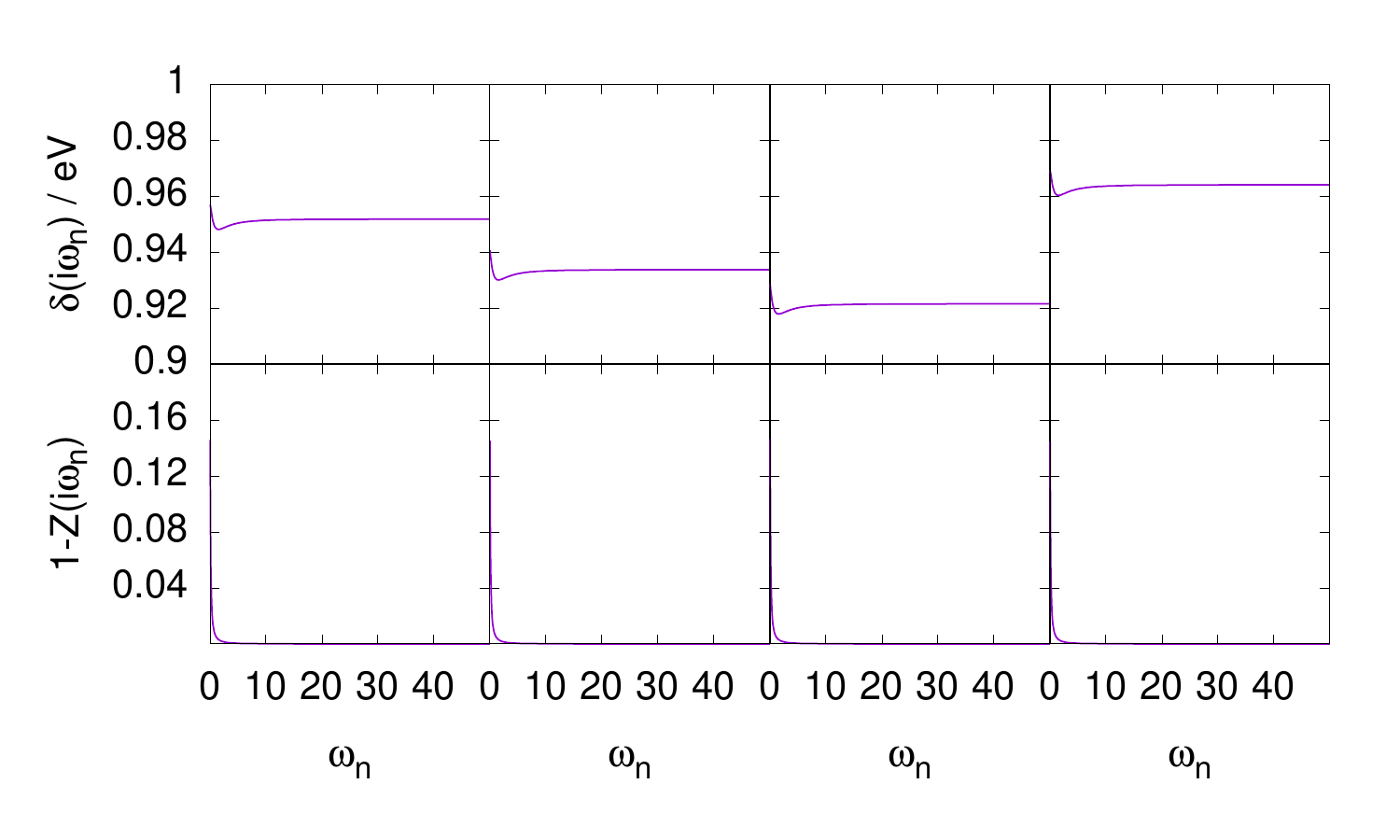}
\caption{(a)-(d) The Matsubara frequency dependence of the gap potential enhancement functions in biased bilayer graphene. (e)-(h) show the associated quasi-particle weight. Each panel shows a different sub-lattice site. Here $\lambda=1$ and $V=\gamma_0$. The bilayer is completely sandwiched so that coupling is with both layers.}
\label{BiasFreqDependence}
\end{figure*}

A total of four on-site potential corrections and their corresponding quasi-particle weights were calculated using Eqns. \ref{Eqn:ABDelta1}-\ref{Eqn:AADelta2}. The Matsubara frequency dependence of these quantities for all the studied lattice
configurations of boron nitride are displayed in Fig.
\ref{Fig:BNmatsubara}. Figures display the frequency dependence at
$\lambda=1$ and a bias potential corresponding to the point of inflection or zero band gap at $V\approx\Delta$. Magnitudes of $\delta(i\omega_n)$
are plotted to aid comparison. For all functions, asymptotic
behaviour is reached at low Matsubara frequencies. \Cref{BiasFreqDependence}
displays the Matsubara frequency dependence of $\delta$ for bilayer
graphene with an electron phonon coupling $\lambda=1$ and bias $V=\gamma_0$. It can be seen
that these functions also quickly settle at a constant asymptotic value. In both cases, the bilayer is completely sandwiched. Results are similar for coupling to the substrate only.

\begin{figure}[!ht]
\centering
\includegraphics[width=0.45\textwidth]{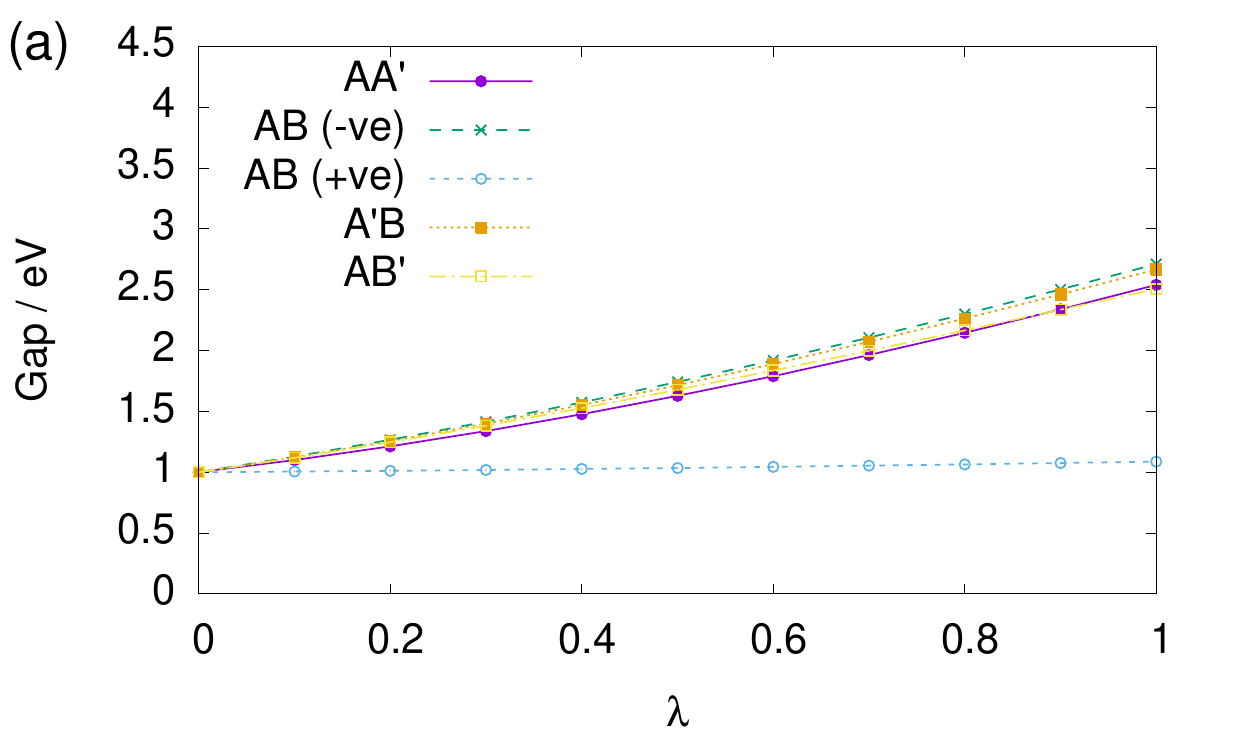}
\includegraphics[width=0.45\textwidth]{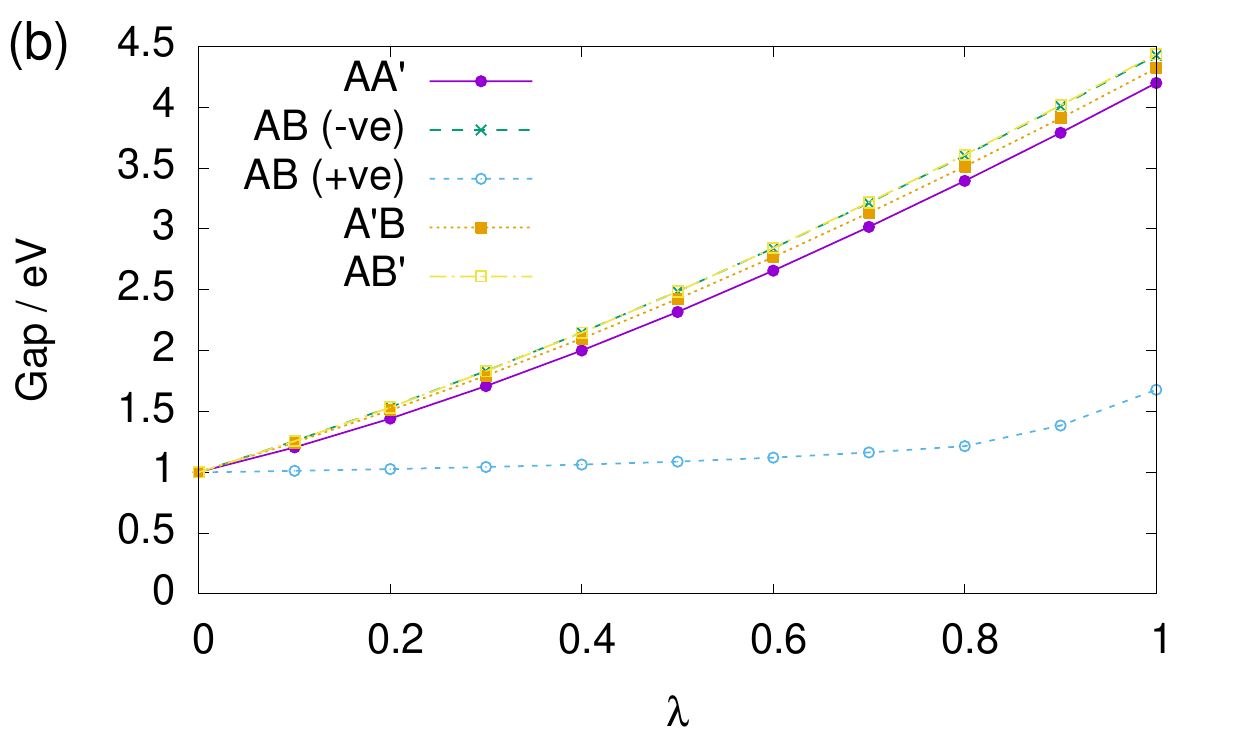}
\caption[Electron band gap evolution with increasing substrate induced electron-phonon interaction strength.]{Electron band gap evolution with increasing substrate induced electron-phonon interaction strength. Panel (a) shows the effect of interaction with a substrate only, and panel (b) with both substrate and superstrate. All stable lattice configurations are set to an original gap size of $1$eV via an applied bias potential before adding the effects of the EPI. The gap associated with applying positive potential difference over the $AB$ configuration is particularly stable against electron-phonon coupling, however the applied potential is large. Other configurations are quite sensitive to EPI.}\label{Fig:OneGap}
\end{figure}

Figure \ref{Fig:OneGap} shows the EPI modified gap for all h-BN stacking
configurations for a non-interacting gap size of $1$eV, and different values
of $\lambda$.  Panel (a) shows the effect of interaction with a substrate only, and panel (b) with both substrate and superstrate, and this convention will be used throughout the remainder of this article. The initial parameters were chosen following Table \ref{Tab:BNBias}
such that the non-interacting tight binding model used for each of the
lattice configurations had a band gap of approximately $1$eV  (where there are 2 values of $V$ leading to a 1eV gap, the smallest $V$ is used). For
increasing electron-phonon coupling, in all cases, the electron band
gap is also increased, and this is typically (although not always) the case, since electron-phonon interactions tend to localise electrons and holes. As the electron-phonon coupling approaches
$\lambda = 1$, the gap modification is quite pronounced. The majority of the lattice structures are highly sensitive to the effects of electron-phonon interaction. The exception is the AB configuration with positive $V$ (i.e. from the bottom to top of the page in Fig. \ref{TBtable}). In that case the gap is quite stable against EPI, although it should be noted that the magnitude of $V$ to obtain a 1eV gap is approximately double in this case. The effect of coupling to both layers rather than to a single layer is that the response of the gap is approximately double that of the bilayers. Otherwise the results are qualitatively similar. This approximate doubling of response is found regardless of bias voltage.

\begin{figure}[!ht]
\centering
\includegraphics[width=0.45\textwidth]{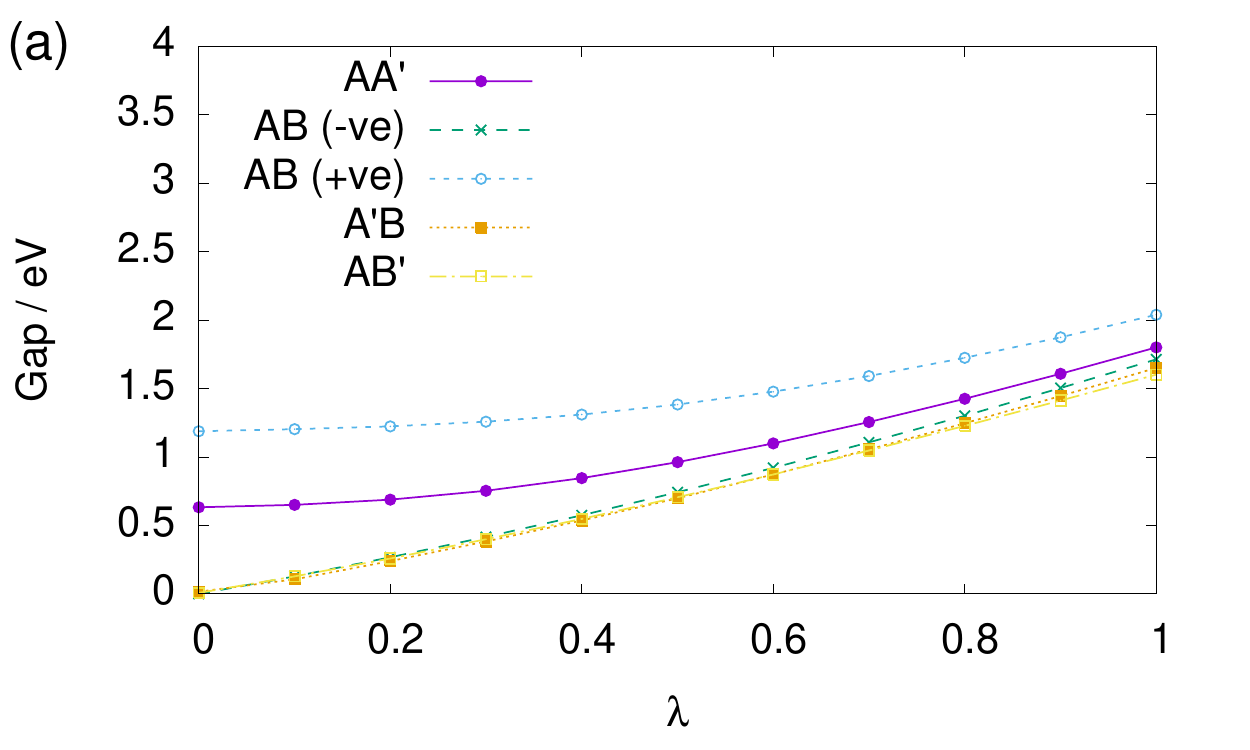}
\includegraphics[width=0.45\textwidth]{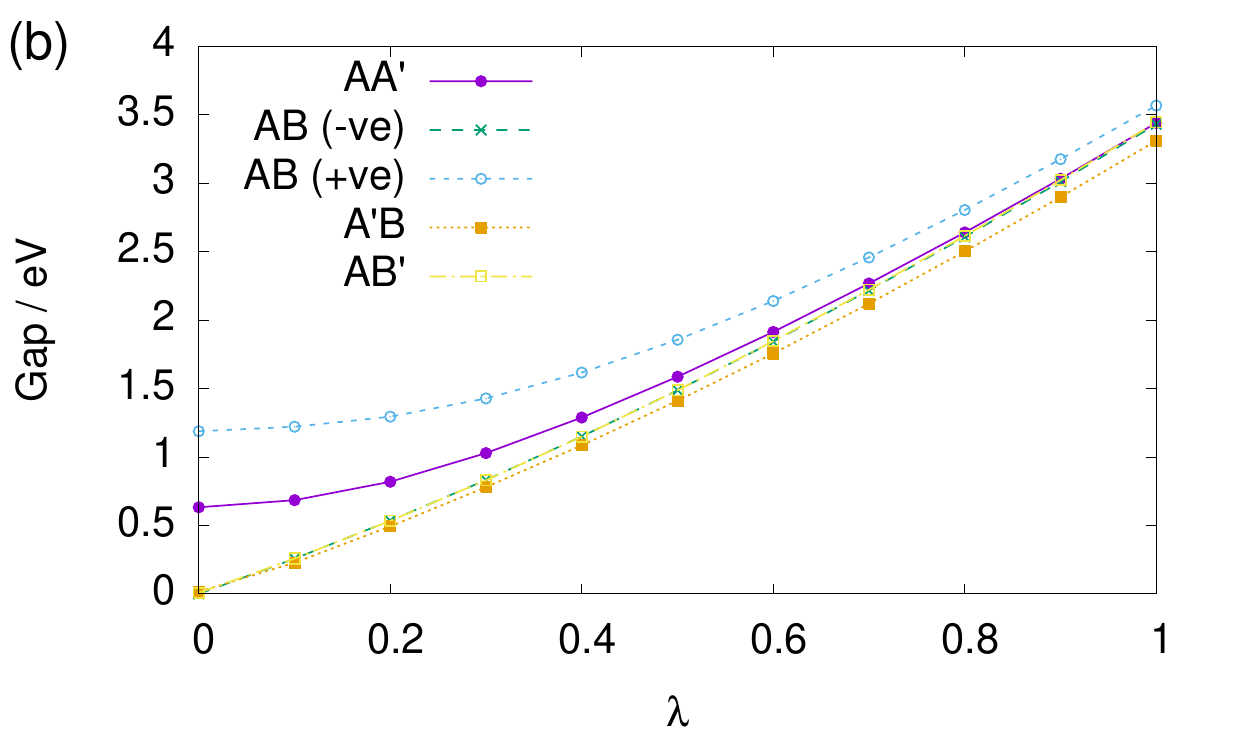}
\caption[Boron nitride electron band gap evolution with increasing electron-phonon interaction strength.]{Electron band gap evolution with increasing electron-phonon interaction strength. $V$ is selected so that the system is at the point of inflection around $V\approx \Delta$ before adding the effects of the electron-phonon interaction. Again, panel (a) shows the effect of interaction with a substrate only, and panel (b) with both substrate and superstrate. Owing to a quadratic response of the gap to the EPI, the $AA'$ configuration and the $AB$ form with positive $V$ are most stable against the effect of interactions for this potential difference between the layers. Zero gap states in the other configurations are not stable against EPI, indicating that attempts to turn off the gap in BN could be sensitive to substrate type.}\label{Fig:MinGapDelta}
\end{figure}

Figure \ref{Fig:MinGapDelta} presents the overall results for band gap evolution when a electron-phonon coupling is added to BN bilayers when $V\approx\Delta$, which is where there is a point of inflection indicating the start of the smaller gradient evolution of the gap on changing $V$ for the $AA'$ and positive $V$ $AB$ configurations, and where there is zero gap for the other configurations. In the case of $AB$ stacked bilayer boron nitride with negative V, $AB'$ and $A'B$ stacking, we see a sharp increase in band gap size similar to that shown in Figure \ref{Fig:OneGap}. This turns on the gap in all 3 cases, and indicates that attempts to switch of the BN gap with potential could be highly sensitive to the strength of electron interactions. On the other hand, in the case of $AA^\prime$ stacking, and $AB$ stacking with positive $V$ a differing situation occurs; for $AA^\prime$ stacking the band gap evolution is much less sensitive to increasing electron-phonon coupling strength. The increase in coupling strength initially has little effect on the band gap, before the band gap evolves towards the same asymptotic behaviour as that seen for the structures with zero gap. Again, this indicates that BN electronics made with $AA'$ stacking and the positive $V$ case of $AB$ stacking could be less sensitive to perturbations from substrates.

\begin{figure}[!ht]
\centering
\includegraphics[width=0.45\textwidth]{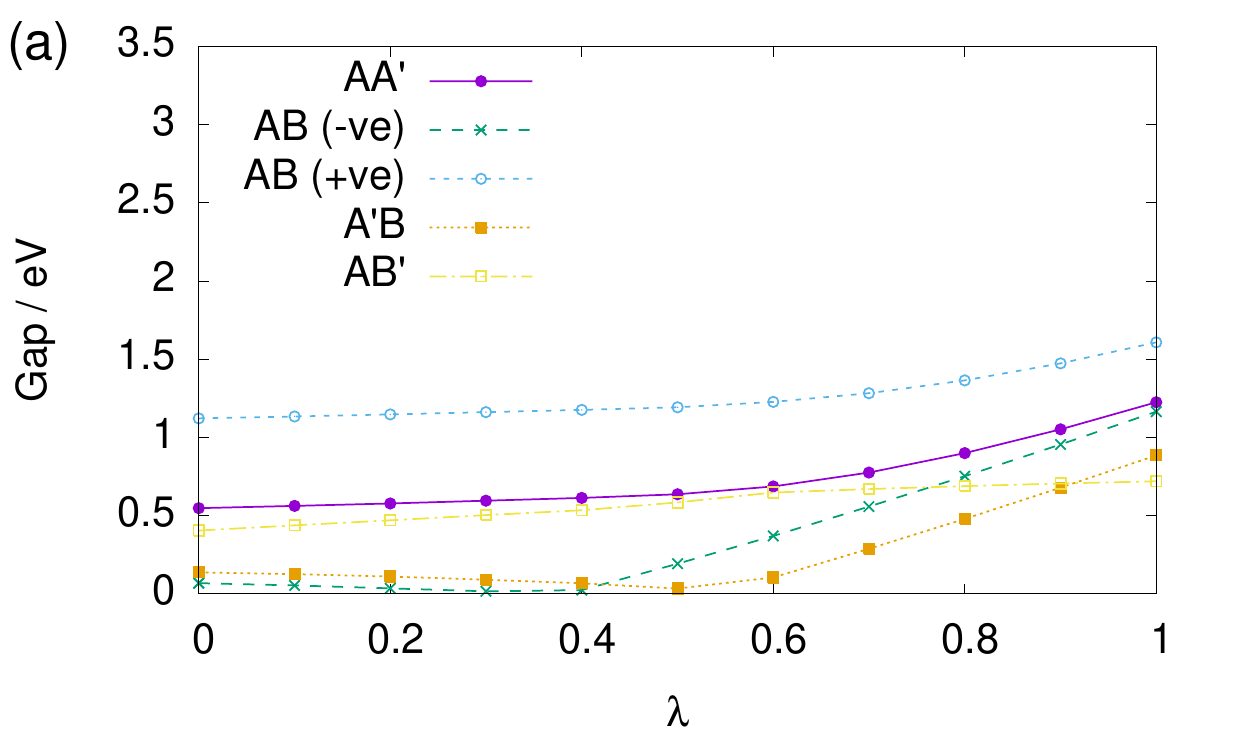}
\includegraphics[width=0.45\textwidth]{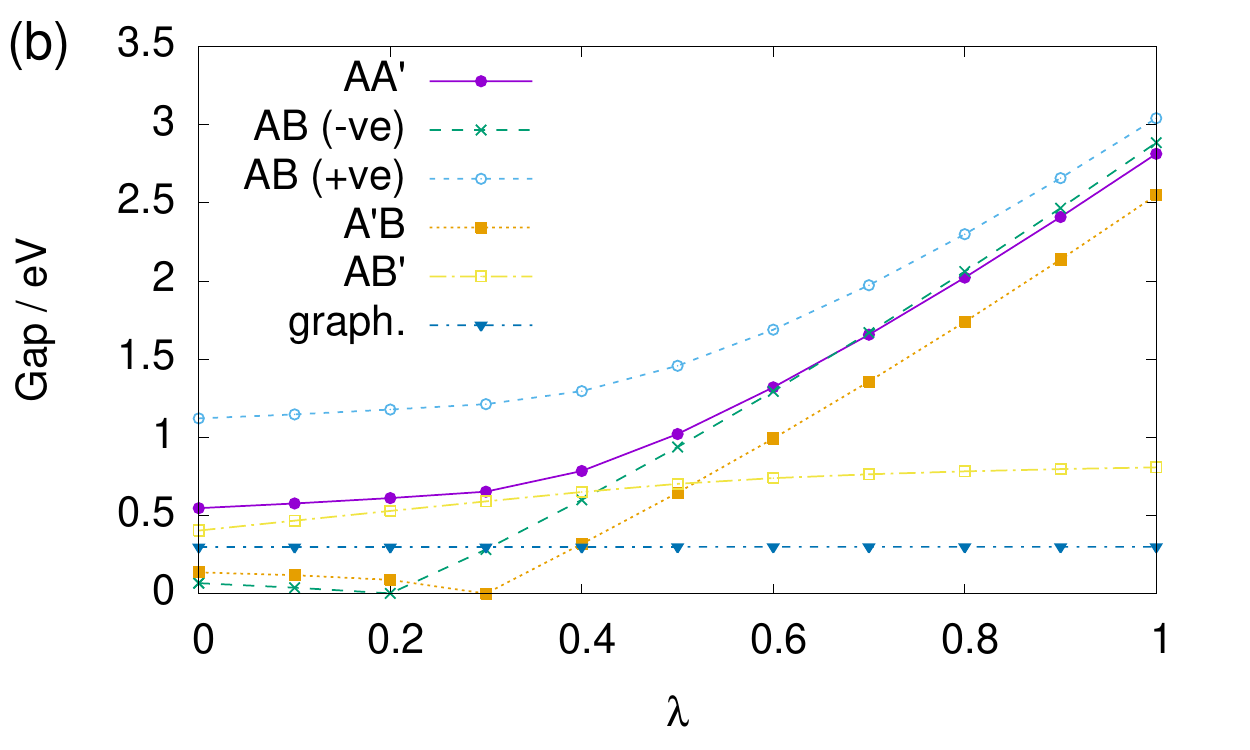}
\caption{Effective potential relative to the applied potential of applied potential $V=\gamma_0$ for various $\lambda$.  Panel (a) shows the effect of interaction with substrate only, and panel (b) with both substrate and superstrate. The graphene gap is stable against EPI, but is relatively small. The gap of the $AB'$ configuration is also relatively stable against EPI, with a gap of around 0.5eV. $AA'$ and $AB$ (+ve $V$) configurations have gaps that are stable for moderate EPI.}
\label{Fig:Gapt}
\end{figure}

\begin{figure}[!ht]
\centering
\includegraphics[width=0.45\textwidth]{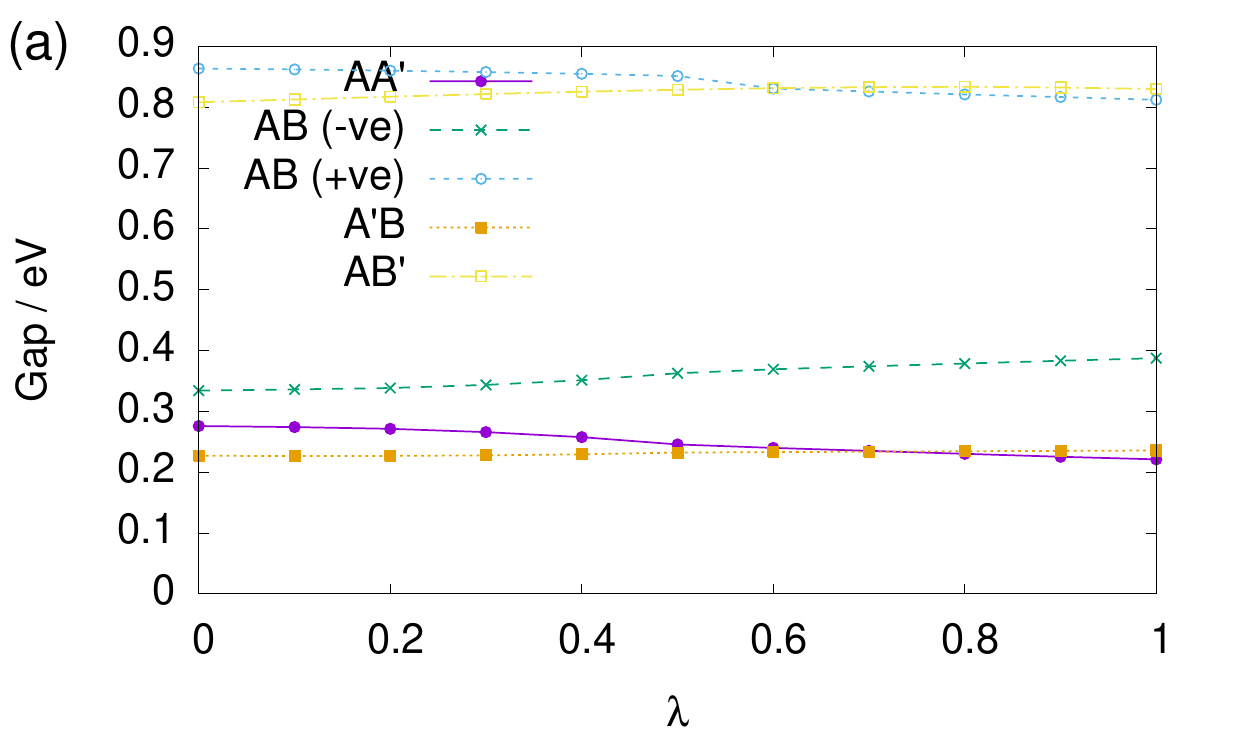}
\includegraphics[width=0.45\textwidth]{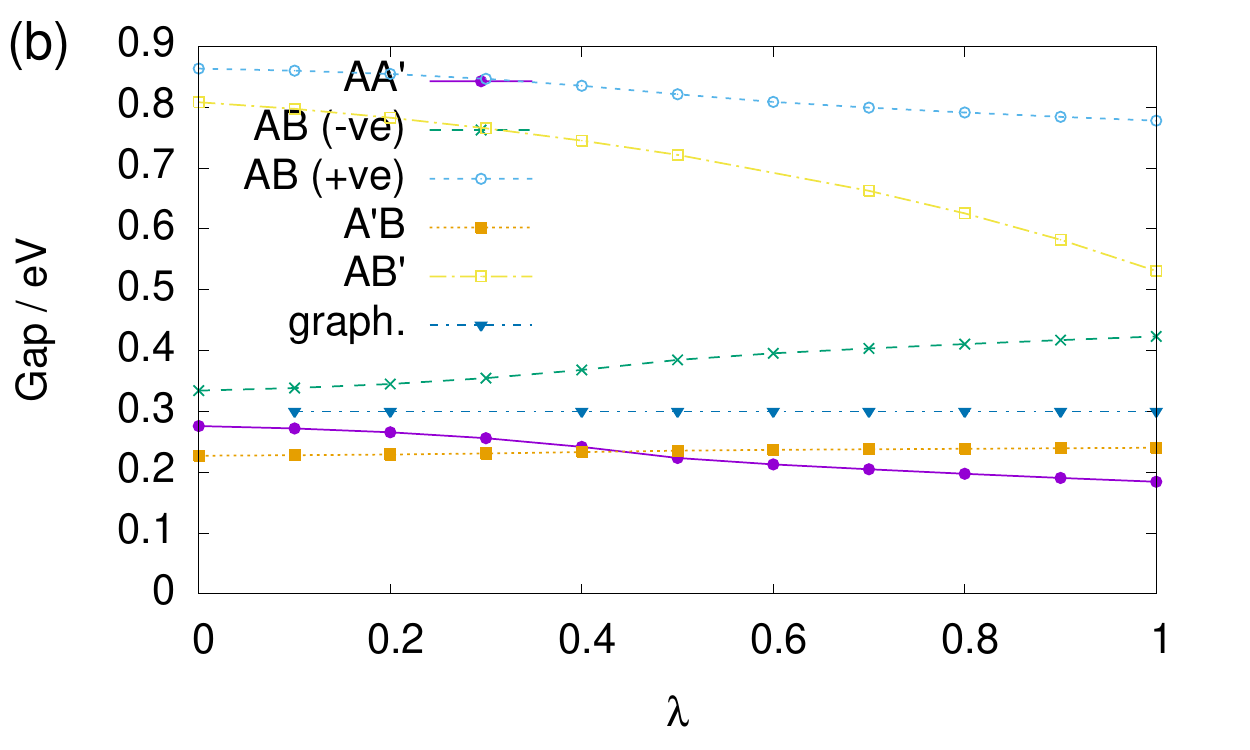}
\caption{Effective potential relative to the applied potential of applied potential $V=2\gamma_0$ for various $\lambda$.  Panel (a) shows the effect of interaction with substrate only, and panel (b) with both substrate and superstrate. For this potential difference between the bilayers, the gap is quite stable over a range of $\lambda$, and there are a wide variety of gap sizes. However, the total potential difference of $2V=4\gamma_0$ is very large at approximately 9.5V.}
\label{Fig:Gaptwot}
\end{figure}

Finally, Figs. \ref{Fig:Gapt} and \ref{Fig:Gaptwot} show evolution of the BN band gaps when $V=\gamma_0$ and $V=2\gamma_0$. These are more stable against electron-phonon interactions than the gaps at $V\approx\Delta$. For comparison, the gap evolution for graphene is shown on both graphs. The graphene gap is highly stable due to the large plateau for applied potentials of up to $V\sim 3\gamma_0$ that can be observed in the non-interacting system as  applied potential is increased. However, the gap size of the biased graphene system is limited to 0.3eV. While the gap in some configurations is unstable to closure followed by gap widening when the electron-phonon interaction is switched on, both the $AA'$ and $AB$ (+ve potential) configurations are very stable against EPI when $V=\gamma_0$, and far more so than for $V\sim\Delta$. Again, this can be related to the broad minimum in the gap seen for the non-interacting model for $AA'$ and $AB$ (+ve $V$) configurations as applied potential is changed (see Fig. \ref{fig:gapvsv}). In particular, only a small percentage change in gap is expected for dimensionless electron-phonon couplings of up to $\lambda\sim 0.3$. The gap in the $AB'$ system seems to be particularly stable against electron interactions for this applied potential. It should therefore be possible to use a wide range of substrates with $AA'$, $AB$ (+ve $V$) and $AB'$ BN configurations without risk of modifying electronic behaviour. Finally, for $V=2\gamma_0$, all structures are reasonably stable against EPI, and a variety of band gaps are available. However, it should be noted that since the total potential difference between the two layers is $2V$, this would correspond to a very large bias between bilayers of approximately 9.5V.


\section{Summary and Conclusion}\label{Sec:Conclusion}

In this paper we have explored the effects of the electron-phonon
interaction on biased graphitic bilayers. Our model is built
specifically using a tight binding parameterisation for BN, which
(with a change of hopping and intrinsic gap size) is applicable for
any ionic graphitic bilayer in any of four possible configurations. We
have also studied biased bilayer graphene in the AB configuration. Perturbative introduction of EPIs to the tight binding model
of these systems was carried out using a Green's function approach. For
graphene and each graphitic bilayer stacking configuration, four equations for
the on-site potential were solved self-consistently. The resulting
on-site potentials were then placed into a tight-binding model and the
effective band gap was calculated. Band gap modification was examined
for four stable and experimentally observable stacking configurations
of BN and one of graphene with varying but significant effects.

Our calculations indicate that when a substrate mediated electron-phonon
interaction is added to biased bilayer graphene, the induced electron
band gap is essentially unmodified once a 300meV gap forms due to a plateau of stability. The choice of
substrate/superstrate is therefore expected to make little difference
to the gap in biased bilayer graphene, which remains relatively small. We note that this is not
contrary to our previous results on gap opening in unbiased graphene
\cite{davenport2014}, rather that the presence of large applied potential difference between layers overwhelms the
effects of the small Coulomb induced inhomogeneity that is responsible for
gap opening in the unbiased case.

On the other hand, for all the stacking configurations, the
band gap in ionic graphitic bilayers has potential to be significantly affected by the electron-phonon
interaction. This modification changes according to the size of the applied potential and the stacking configuration. While it is in principle possible to reduce the gap to zero to induce graphene like properties  in certain stacking configurations of bilayer graphitic materials, our calculations indicate that interaction with substrates could mean that different applied potentials would be needed for different substrates, making such devices difficult to tune. Ionic bilayers with both an $AA'$ configuration and an $AB$ configuration and positive applied bias each have a broad minimum in gap size that extends from around $V\approx\Delta$ to $V\approx 3\gamma_0$. Electron-phonon interactions change the effective size of the applied potential, and therefore any near constant region in the response of the gap to change of applied potential can lead to a region of relative stability against electron-phonon interactions, and therefore better compatibility between devices made on different substrates. Response to interactions in all configurations tends to decrease steadily once $V>\Delta$, with the response becoming steadily flatter until $V\sim 3\gamma_0$.

There are several limitations to the calculations presented here. The most significant is the mean-field (local) approach to self consistency. We also note that the lowest order perturbation theory is used. While use of the local approach is quite standard for this type of calculation, one can go beyond the mean field theory, for example by using the dynamical cluster approximation. Detailed DCA calculations using higher order perturbation theory have been carried out for monolayers of graphene on substrates in Ref. \cite{Hague2014} showing only quantitative differences with the mean-field results. We would expect similar quantitative differences in the results for the bilayer systems discussed here, but nothing qualitative. The quantitative changes are a reduction in the gap modification when non-local fluctuations are present, and an increase in the gap modification when higher order terms in the perturbation theory are introduced. We suggest this type of calculation to be the next logical step, although such calculations would be very complex for bilayers and we would not expect any significant changes to the conclusions. We also note that the EPIs are simplified by the rigidity introduced by the substrate. Additional flexural contributions could lead to modulation of the hopping terms in the case of freestanding bilayers, although such terms would  be smaller than those in monolayers due to the extra rigidity of the bilayer. 

This article has focused on using tight binding parameterisations of
BN as a starting point for the calculations. In fact, the results will
be valid for a range of other graphitic bilayers such as ZnO,GaN, AlN,
BeO and MgO, although there will be quantitative differences in the
size of interlayer hopping and the intrinsic gap. The main difference
being that the induced gap will be different, especially that associated with
the plateaus, which are sensitive to the interlayer hopping. Also,
results are expected to be qualitatively similar for metal
dichalcogenides (noting that the metal dichalcogenides have 6 atoms
per unit cell, which would increase the complexity of the
calculations). Such starting materials typically have lower bare gaps
than BN, so the regions where gaps are stable against interactions
would be reached with lower potential differences across the
planes. Results presented here may be particularly important for
optimising devices that use strongly polarisable substrates.  Some of
these materials are already being used in experimental electronic
devices, and MoS$_2$ has been the subject of a lot of recent interest
for its use in FETs. It would not be surprising to see other graphitic
bilayers used for such a purpose in the near future.

\section*{Acknowledgements}
JPH would like to acknowledge EPSRC grant No. EP/H015655/1.

\bibliography{Bilayer_BoronNitride}

\end{document}